\documentclass[12pt]{iopart}
\usepackage{graphicx}
\usepackage{graphics}

\begin{document}

\title[Power-law behavior and condensation phenomena in disordered urn models]
{Power-law behavior and condensation phenomena in disordered urn models}

\author{Jun-ichi Inoue$^1$ and Jun Ohkubo$^2$}
\address{$^1$Complex Systems Engineering, 
Graduate School of Information Science and 
Technology, Hokkaido University, 
N14-W9, Kita-ku, Sapporo 060-0814, Japan}

\address{$^2$Institute for Solid State 
Physics, University of Tokyo, 
Kashiwanoha 5-1-5, Kashiwa, Chiba 
277-8581, Japan}

\eads{$^1$j$\underline{\,\,\,}$inoue@complex.eng.hokudai.ac.jp, 
$^2$ohkubo@issp.u-tokyo.ac.jp}

%%%%%%%%%%%%%%%%%%%%%%%%%%%%%%%%%%%%%%%%%%%%%%%%%%%%%%%%%%%%%%%%%%%
%%                   Abstract                                   %%%
%%%%%%%%%%%%%%%%%%%%%%%%%%%%%%%%%%%%%%%%%%%%%%%%%%%%%%%%%%%%%%%%%%% 
\begin{abstract}
We investigate equilibrium statistical properties 
of urn models with disorder. Two urn models 
are proposed; one belongs to the Ehrenfest class,
and the other corresponds to the Monkey class.
These models are introduced from the view point 
of the power-law behavior and randomness; it is clarified that 
quenched random parameters play an important role 
in generating power-law behavior. We evaluate the occupation probability 
$P(k)$ with which an urn has $k$ balls by using the concept of 
statistical physics of disordered systems. 
In the disordered urn model belonging to the Monkey class, 
we find that above critical density $\rho_\mathrm{c}$ 
for a given temperature, condensation phenomenon occurs 
and the occupation probability changes its scaling 
behavior from an exponential-law to a heavy 
tailed power-law in large $k$ regime. 
We also discuss an interpretation of our results 
for explaining of macro-economy, in particular, 
emergence of wealth differentials. 
\end{abstract}
\submitto{\JPA}
\maketitle

%%%%%%%%%%%%%%%%%%%%%%%%%%%%%%%%%%%%%%%%%%%%%%%%%%%%%%%%%%%%%%%%%%%%%%%%%%
\section{Introduction}
\label{sec:Intro}
%%%%%%%%%%%%%%%%%%%%%%%%%%%%%%%%%%%%%%%%%%%%%%%%%%%%%%%%%%%%%%%%%%%%%%%%%%
A lot of techniques and concepts of statistical 
mechanics of disordered spin systems, in particular, 
the replica method originally used to analyze 
the thermodynamics of spin glass model 
by Sherrington and Kirkpatrick \cite{SK}, 
have been applied to various research fields 
beyond conventional 
physics, i.e. information processing \cite{Nishi}, 
game theory \cite{Coolen} and so on. 
The exactly solvable mathematical model, 
which describes these problems, 
is categorized as mean-field class \cite{Mezard}. 

On the other hand, as another exactly tractable model, 
in 1907, Paul and
Tatiana Ehrenfest published a paper corroborating
Boltzmann's view of thermodynamics \cite{Ehrenfest}. 
Their urn model has been defined by Kac \cite{Kac} as an 
exactly solvable example in statistical physics.  
While it has also been criticized as 
a marvelous exercise too far removed from reality,
their urn model has been applied to modern problems such as 
complex networks \cite{Adamic, Barabasi} or econophysics 
\cite{Mantegna2000,Bouchaud2000}, etc. 
For instance, 
based on extensive simulations of 
the Lennard-Jones fluid
requiring in part a parallel 
computer in Juelich, an Italian-German team
has shown that the prediction of the Ehrenfest 
urn effectively describes
the behavior of the gas phase \cite{Scalas}. 
Moreover, it has been revealed that 
the mathematical structure of equilibrium 
state of the urn model \cite{Luck} 
is similar to the zero-range process,
which has been widely investigated in research fields 
of non-equilibrium statistical physics \cite{Evans}.

Recently, 
in the research field of 
complex networks \cite{Adamic,Barabasi}, 
Ohkubo et al. \cite{Ohkubo} 
proposed a network model based on `Ehrenfest class urn model' 
to explain how the complex network gets scale-free-like properties,
where `Ehrenfest class' means that each urn has distinguishable balls. 
In the model, 
each urn corresponds to a 
node in graph (network) and 
the number of 
distinguishable 
balls, $k$, in each urn is regarded as 
degree of nodes. 
For this model system, 
they succeeded in 
deriving 
the scale-free-like properties 
$\sim k^{-2}(\log k)^{-2}$ 
in the probability of the degree of nodes by the usage of
the replica symmetric theory \cite{Ohkubo2006}. 
In addition, 
the similarity between the disordered urn model
and the random field Ising model \cite{Ohkubo2007_1}, 
and the condensation phenomena in the disordered 
urn model have been investigated \cite{Ohkubo2007_2}.

We here note that 
there are a lot of works in which 
the power-law behavior and the condensation phenomena
in urn models have been studied \cite{Luck,Bialas}.
For example, in the zeta urn model \cite{Luck}, 
the power-law behavior in the probability of the number of balls,
i.e., the occupation distribution,
stems from a power-law form of the Boltzmann weight.
However,
when we attempt to describe various problems in the real world, 
we should take into account the disorder and 
treat the urns as a heterogeneous system.
Mainly, the previous models which cause power-law behavior
in the occupation distribution
do not contain any disorder,
and hence it would be important to investigate
`disordered' urn models which cause the power-law distribution.

In this paper, we propose two disordered urn models in which quenched
randomness is important for generating the 
power-law behavior. One of them belongs 
to the Ehrenfest class, and the other corresponds
to the Monkey class, in which each urn has indistinguishable balls. 
In particular, for the Monkey class urn model,
we investigate a real-space 
condensation phenomenon 
in which macroscopic number 
of balls are condensed into only one urn. 
The occupation probability $P(k)$ with 
which an urn has $k$ balls
is calculated analytically,
and furthermore, the critical density $\rho_{\rm c}$ for a given temperature
is evaluated. As the result, 
it is shown that the occupation distribution 
function $P(k)$ changes its scaling behavior 
from the exponential $k^{-(\alpha +1)}\,{\rm e}^{-k}$-law 
to the $k^{-(\alpha +2)}$ power-law in large $k$ regime.

This paper is organized as follows. In the 
next section \ref{sec:General}, 
we introduce the general formalism for the urn model 
with an arbitrary energy function. 
Although there are several analytical treatments for disordered
urn models \cite{Ohkubo2006,Leuzzi}, 
we give the formalism for the disordered
urn model with an arbitrary energy function with a different
point of view, and additionally, the analytical treatment
makes this paper self-contained.
This analytical treatment contains 
both Ehrenfest and Monkey classes as its special cases. 
We explain the relation between the saddle point, that 
determines the thermodynamic properties of the system,
and the chemical potential. With the assistance of 
this general formalism, we provide an analysis for a special 
choice of the energy function, which is an example of 
the Ehrenfest class in section \ref{sec:Ehrenfest}. 
We discuss the condition on which the power-law appears 
in the tail of the occupation probability for the model. 
In section \ref{sec:BE}, we show that 
the condensation occurs 
for the special case of the Monkey class with disorder,
and a heavy tailed power-law 
emerges in the occupation probability. 
In section \ref{sec:Econophysics}, 
we provide a possible link 
between our results and macro economy, in particular, 
wealth differentials. 
Last section is a summary.  

%%%%%%%%%%%%%%%%%%%%%%%%%%%%%%%%%%%%%%%%%%%%%%%%%%%%%%%%%%%%%%%%%%%%%%%%
\section{General formalism for urn models with disorder}
\label{sec:General}
%%%%%%%%%%%%%%%%%%%%%%%%%%%%%%%%%%%%%%%%%%%%%%%%%%%%%%%%%%%%%%%%%%%%%%%%
%%We first formulate the general class of the urn model which 
%%contains both Ehrenfest and Monkey classes as its two special cases. 
%%%%%%%%%%%%%%%%%%%%%%%%%%%%%%%%%%%%%%%%%%%%%%%%%%%%%%%%%%%%%%%%%%%%%%%%
Let us prepare $N$ urns and $M$ balls ($M \equiv \rho N$) and 
consider the situation in which the $N$ urns share the $M$ balls.
Then, we start our argument from the Ehrenfest class 
urn model \cite{Luck} in which each ball in urns is distinguishable. 
For the mathematical model categorized in the Ehrenfest class, 
the Boltzmann weight $p (n_{i})$ that $i$-th urn 
possesses $n_{i}$ balls is given by  
%%%%%%%%%%%%%%%
\begin{eqnarray}
p(n_{i}) & = & 
\frac{{\exp}
\left[
-\beta E(\epsilon_i, n_{i})
\right]}{n_{i}!},
\end{eqnarray}
%%%%%%%%%%%%%%%
where $E(\epsilon_i, n_i)$ is an energy function,
$\epsilon_i$ a disorder parameter for urn $i$,
and $\beta$ the inverse temperature of the system.
The factorial $n!$ stems from the property of the Ehrenfest class \cite{Luck}.
%%%%%%%%%%%%%%%%%%%%%%%%%%%%%%%%%%%%%%%%%%%%%%%%%%%%%%%%
The only point of our analysis which is different 
from \cite{Luck} is quenched disorder $\epsilon_{i}$ 
appearing in the energy function. 
%%%%%%%%%%%%%%%%%%%%%%%%%%%%%%%%%%%%%%%%%%%%%%%%%%%%%%%%%
For the Ehrenfest class, the probability that an urn specified 
by the disorder parameter $\epsilon_1$ possesses $k$ balls is given by
%%%%%%%%%%%%%
\begin{eqnarray}
\fl
f_{k}^{(\epsilon_1)} (\{\epsilon_{i/1}\}) = 
\frac{1}{Z_{1}}\,
\sum_{n_{1}=0}^{\infty}
\cdots
\sum_{n_{N}=0}^{\infty}
\delta (n_{1},k)\,
p(n_{1}) 
\cdots
p(n_{N})
\,\delta 
\left(
\sum_{i=1}^{N}n_{i},M
\right) \nonumber \\
\fl
\mbox{} =  
\frac{e^{-\beta E(\epsilon_1, k)}}{k!}
\frac{1}{Z_{1}}
\prod_{i=2}^{N}
\left[ 
\sum_{n_{i}=0}^{\infty}
\left(
\frac{{\rm e}^{-\beta E(\epsilon_i, n_{i})}}
{n_{i}!}
\right)
\right]
\oint 
\frac{dz}{2\pi i}
{\exp}
\left[
\left(k+\sum_{i=2}^{N}
n_{i}-M-1
\right)
\log z
\right] \nonumber \\
\fl 
\mbox{} = 
\frac{e^{-\beta E(\epsilon_1, k)}}{k!}
\left(
\frac{Z_{2}}{Z_{1}}
\right)
\label{eq:Ehren},
\end{eqnarray}
%%%%%%%%%%%
where $\{\epsilon_{i/1}\} 
\equiv \{\epsilon_{2},\cdots,\epsilon_{N}\}$ 
and we defined $Z_{2}$ as 
%%%%%%%%%%%%%%
\begin{eqnarray}
\fl
Z_{2} =  
\prod_{i=2}^{N}
\left[
\sum_{n_{i}=0}^{\infty}
\left(
\frac{{\rm e}^{-\beta E(\epsilon_i, n_{i})}}
{n_{i}!}
\right)
\right] 
\oint 
\frac{dz}{2\pi i}
{\exp}
\left[
\left(
k+\sum_{i=2}^{N}
n_{i}-M-1
\right) \log z
\right],
\end{eqnarray}
%%%%%%%%%
and used the Fourier 
transform of 
the Kronecker-delta: 
%%%%%%%%%%%%%%%%%%
\begin{eqnarray}
\delta (A,B ) & = & 
\oint 
\frac{dz}{2\pi i}
\,z^{A-B-1}
\end{eqnarray}
%%%%%%%%%%%%%%%%%%%%%%
to introduce 
the conservation 
of the total balls : 
$n_{1}+\cdots +n_{N}=M \equiv \rho N$ 
into the system.
In order to calculate $P(k)$,
we take the configuration 
average of $f_k^{(\epsilon_1)}(\{\epsilon_{i/1}\})$.
While one can calculate the configuration average 
by means of the replica method \cite{Ohkubo2006},
it has been revealed that the mathematical structure of 
the disordered urn model
is related to that of a random field Ising model \cite{Ohkubo2007_1}.
Hence, we here use the law of large numbers and simplify the calculation.
%%%%%%%%%%%%%%%%%%%%%%%%%%%%%%%%%%%%%%%%%%%%%%%%%%%%%%%%%%%%%%%%%%%%%%%%%%
To calculate the average of the 
quantity $\exp [\log f_{k}^{(\epsilon_{1})}
(\{\epsilon_{i/1}\})]$ over the configuration 
$\{\epsilon_{1},\cdots,\epsilon_{N}\}$, 
we consider the Tayler-expansion: 
%%%%%
\begin{eqnarray}
\fl
\langle 
\exp 
[\log f_{k}^{(\epsilon_{1})}
(\{\epsilon_{i/1}\})]
\rangle_{\{1,2,\cdots,N\}} & = & 
1 + 
\langle 
\log f_{k}^{(\epsilon_{1})}
(\{\epsilon_{i/1}\})
\rangle_{\{1,2,\cdots,N\}} \nonumber \\
\mbox{} & + & 
\frac{1}{2}
\langle 
(\log f_{k}^{(\epsilon_{1})}
(\{\epsilon_{i/1}\}))^{2}
\rangle_{\{1,2,\cdots,N\}}+\cdots 
\label{eq:tayler1}
\end{eqnarray}
%%%%%%%%%%
where $\langle \cdots \rangle_{\{1,2,\cdots,N\}}$ 
means the configuration average
over $\{ \epsilon_1, \cdots,  \epsilon_N\}$.
%%%%%%%%%
Here we assume that the observable 
$\log 
f_{k}^{(\epsilon_{1})}
(\{\epsilon_{i/1}\})$ 
for a given realization of 
configuration 
$\{\epsilon_{1},\cdots,\epsilon_{N}\}$ 
is almost identical to the average 
$\langle \log 
f_{k}^{(\epsilon_{1})}
(\{\epsilon_{i/1}\}) \rangle_{\{1,2,\cdots,N\}}$. 
In other words, 
the deviation is vanishingly small as 
$\langle 
(\log f_{k}^{(\epsilon_{1})}
(\{\epsilon_{i/1}\}))^{2}
\rangle_{\{1,2,\cdots,N\}}
-
\langle 
\log f_{k}^{(\epsilon_{1})}
(\{\epsilon_{i/1}\})
\rangle_{\{1,2,\cdots,N\}}^{2} \simeq 0$ 
in the thermodynamics limit. 
By using the assumption, 
we rewrite (\ref{eq:tayler1}) as follows. 
%%%%%
\begin{eqnarray}
\fl
\langle 
\exp 
[\log f_{k}^{(\epsilon_{1})}
(\{\epsilon_{i/1}\})]
\rangle_{\{1,2,\cdots,N\}} & \simeq & 
1 + 
\langle 
\log f_{k}^{(\epsilon_{1})}
(\{\epsilon_{i/1}\})
\rangle_{\{1,2,\cdots,N\}} + 
\frac{1}{2}
\langle \log f_{k}^{(\epsilon_{1})}
(\{\epsilon_{i/1}\})
\rangle_{\{1,2,\cdots,N\}}^{2} \nonumber \\
\mbox{} & + & \cdots =  
\exp
[
\langle 
\log  
f_{k}^{(\epsilon_{1})}
(\{\epsilon_{i/1}\})
\rangle_{\{1,2,\cdots,N\}}]
\label{eq:tayler2}
\end{eqnarray}
%%%%%%%%%%%
This replacement 
of the configuration average 
reads 
%%%%%%
\begin{eqnarray}
\fl
\langle f_k^{(\epsilon_1)} 
(\{\epsilon_{i/1}\})
\rangle_{\{1,2,\cdots,N\}}= 
\left\langle \exp 
\left(  \log f_k^{(\epsilon_1)}  
(\{\epsilon_{i/1}\})
\right)  \right\rangle_{\{1,2,\cdots,N \}} \simeq 
\exp \left( \langle  \log  f_k^{(\epsilon_1)} 
(\{\epsilon_{i/1}\})
\rangle_{\{1,2,\cdots,N\}} \right) \nonumber \\
\fl
= \exp \left( \left\langle 
\log [ e^{-\beta E(\epsilon_1,k)} / k!] 
+ \langle \log Z_2 \rangle_{\{ 2,3,\cdots,N\}} - \langle \log Z_1 \rangle_{\{2,3,\cdots,N\}} 
\right\rangle_{\{1\}} \right). 
\end{eqnarray}
%%%%%%%%%%%%%%%%%%%%%%%%
%%%%%%%%%%%%%%%%%%%%%%%
In the thermodynamic limit 
$N \to \infty$, 
$\langle \log Z_{2} \rangle_{\{2,3,\cdots,N\}}$ is evaluated as 
%%%%%%%%%%%%%%%%%%
\begin{eqnarray}
\fl
\langle \log Z_{2} \rangle_{\{2,3,\cdots,N\}} 
= 
\left\langle 
\log \left( \oint 
\frac{dz}{2\pi i}
z^{k-M-1}
\prod_{i=2}^N \left\{
\sum_{n_i=0}^{\infty}
\frac{{\rm e}^{-\beta E(\epsilon_i, n_i)}}
{n_i!}
z^{n}
\right\} \right)
\right\rangle_{\{2,3,\cdots,N\}} \nonumber \\
\fl 
= 
\left\langle 
\log \left( \oint 
\frac{dz}{2\pi i}
\exp\left[ (k-\rho N-1) \log z
+ \sum_{i=2}^N \left\{
\log \sum_{n_i=0}^{\infty}
\frac{{\rm e}^{-\beta E(\epsilon_i,n_i)}}
{n_i!}
z^{n}
\right\} \right] \right)
\right\rangle_{\{2,3,\cdots,N\}} \nonumber \\
\fl 
\simeq 
\log \left( \oint 
\frac{dz}{2\pi i}
\exp\left[ (k-\rho N-1) \log z
+ 
(N-1)\left\langle \log \sum_{n=0}^{\infty}
\frac{{\rm e}^{-\beta E(\epsilon,n)}}
{n!}
z^{n} \right\rangle
\right] \right) \nonumber \\
\fl
\simeq
k\log z_{s1} -
(\rho N +1) 
\log z_{s1}
+(N-1) {\biggr \langle} \log 
\sum_{n=0}^{\infty}
\frac{{\rm e}^{-\beta E(\epsilon,n)}}
{n!}z_{s1}^{n}
{\biggr \rangle}
\label{eq:geneZ2},
\end{eqnarray}
%%%%%%%%%%%%%%%%%%%%%%
where we used the law of large numbers,
and in the final line the saddle point method was used.
$\langle \cdots \rangle$ means the average over only $\epsilon$, 
namely, $\langle \cdots \rangle \equiv 
\int (\cdots) D(\epsilon) d\epsilon$. 
%%%%%%%%%%%%%%%%%%%%%%%%%%%%%%%%%%
In the next two sections, 
we consider specific choices of $D(\epsilon)$ 
to evaluate the occupation probability 
distribution concretely. 
Using the same way as the $Z_{2}$, $Z_{1}$, which 
is obtained by the normalization condition of $f_{k}^{(\epsilon_1)}$, 
namely, $\sum_{k=0}^{\infty}f_{k}^{(\epsilon_1)}=1$ with 
equation (\ref{eq:Ehren}), is rewritten as
%%%%%%%%%%%%%%%%
\begin{eqnarray}
\fl
\langle \log Z_{1} \rangle_{\{2,3,\cdots,N\}} \nonumber \\
\fl
= 
\log 
\sum_{k=0}^{\infty}
\frac{{\rm e}^{-\beta E(\epsilon_1,k)}}
{k!}z_{s2}^{k} -
(\rho N +1)
\log z_{s2} +
(N-1)
{\biggr \langle}
\log 
\sum_{n=0}^{\infty}
\frac{{\rm e}^{-\beta E(\epsilon,n)}}
{n!}
z_{s2}^{n}
{\biggr \rangle}
\label{eq:geneZ1}.
\end{eqnarray}
%%%%%%%%%%%%%%%%
We easily find 
$z_{1s} = z_{2s}$ 
because the first terms for 
each saddle point equation 
(\ref{eq:geneZ2}) or (\ref{eq:geneZ1}) are 
vanishingly smaller than the other two terms 
in the limit of $N \to \infty$. 

Thus, we obtain the saddle point 
equation with respect to $z_{s} \equiv z_{s1} = z_{s2}$ and 
the occupation probability 
$P(k)= \langle f_{k}^{(\epsilon)} \rangle$
that an arbitrary 
urn with the energy function $E$ at inverse temperature $\beta$ 
has $k$ balls are given by 
%%%%%%%%%%%%%%%%%%%%%%%
%%%%%%%%%%%%%%%%%%%%
\begin{eqnarray}
\rho z_{s}^{-1} & = & 
{\biggr \langle}
\frac{\sum_{n=0}^{\infty} n
\frac{{\rm e}^{-\beta E(\epsilon,n)}}
{(n-1)!}z_{s}^{n-1}}
{
\sum_{n=0}^{\infty}
\frac{{\rm e}^{-\beta E(\epsilon,n)}}
{n!}z_{s}^{n}}
{\biggr \rangle}
\label{eq:saddle}
\end{eqnarray} 
%%%%%%%%%%%%%
and
%%%%%%%%%%%%%%%%
\begin{eqnarray}
P (k) & = & 
{\biggr \langle}
\frac{
\frac{{\rm e}^{-\beta E(\epsilon,k)}}
{k!}z_{s}^{k}}
{
\sum_{n=0}^{\infty}
\frac{{\rm e}^{-\beta E(\epsilon,n)}}
{n!}z_{s}^{n}}
{\biggr \rangle}, 
\label{eq:dist}
\end{eqnarray}
%%%%%%%%%%%%%%%%%
respectively. 
It should be noticed that 
the above saddle point equation for the 
Ehrenfest 
class urn model 
(\ref{eq:saddle}) is now 
rewritten in terms of 
chemical potential 
%%%%%%%%%%
\begin{eqnarray}
\mu & \equiv & 
\beta^{-1}
\log z_{s}
\label{eq:chemi}
\end{eqnarray}
%%%%%%%%%%
as 
%%%%%%%%
\begin{eqnarray}
\rho & = & 
{\biggr \langle}
\frac{
\sum_{n=0}^{\infty}
n 
\frac{{\rm e}^{-\beta [E(\epsilon,n)-n\mu]}}
{n!}}
{
\sum_{n=0}^{\infty}
\frac{{\rm e}^{-\beta [E(\epsilon,n)-n\mu]}}
{n!}
}
{\biggr \rangle}.
\label{eq:def_rho}
\end{eqnarray}
%%%%%%%%%%%%%%%%%%%%%%%%
Then, 
we define 
the probability 
$p_{n}$ that 
an arbitrary Ehrenfest class urn with 
energy $E$ has $n$ balls by 
%%%%%%%%%%%%%%%%%%%
\begin{eqnarray}
p_{n} & = & 
\frac{\phi_{E,\mu,\beta} (\epsilon,n)}
{\sum_{n=0}^{\infty}
\phi_{E,\mu,\beta} (\epsilon,n)},\,\,\,\,\,
\phi_{E,\mu,\beta} (\epsilon, n) =
\frac{{\rm e}^{-\beta [E(\epsilon,n)-n\mu]}}
{n!}.
\label{eq:dist_n}
\end{eqnarray}
%%%%%%%%%%%%%%%%%%%%%%%%%%%%%
From this formula of the probability $p_{n}$ 
with the effective Boltzmann factor 
$\phi_{E,\mu,\beta}(\epsilon,n)$, the equation 
(\ref{eq:def_rho}) means that 
the ratio $M/N$ corresponds 
to the average number of 
balls put in an arbitrary 
urn: 
$\rho =\sum_{n=0}^{\infty}
n \phi_{E,\mu,\beta}(\epsilon,n)/\sum_{n=0}^{\infty}
\phi_{E,\mu,\beta}(\epsilon,n)$, and 
its value is 
controlled by 
the chemical potential $\mu$ 
through the equation (\ref{eq:def_rho}). 
Then, the chemical potential $\mu$ and 
the saddle point $z_{s}$ 
are related through the equation 
(\ref{eq:chemi}). 
Therefore, 
when we construct 
the system so as to have a 
density $\rho$, the corresponding 
saddle point $z_{s}$ is given by (\ref{eq:def_rho}). 
As the result, the chemical 
potential $\mu$ that gives $\rho$ is 
determined by the relation (\ref{eq:chemi}). 

Thus, our problem is now to solve 
the saddle point equation 
%%%%%%%%%
\begin{eqnarray}
\rho & = & 
{\biggr \langle}
\frac{\sum_{n=0}^{\infty}
n\, \phi_{E,\mu,\beta} (\epsilon,n)}
{\sum_{n=0}^{\infty}
\phi_{E,\mu,\beta} (\epsilon,n)}
{\biggr \rangle}
\label{eq:gene_saddle},
\end{eqnarray}
%%%%%%
and 
to calculate 
the following averaged occupation probability
for the solution 
$z_{s}={\rm e}^{\beta \mu}$
of the equation (\ref{eq:gene_saddle}):
%%%%%%%%%%%%%%%%%%%%%%%
%%%%%%%%%%
\begin{eqnarray}
P (k) & = & 
{\biggr \langle} 
\frac{
\phi_{E,\mu,\beta} (k)}
{\sum_{n=0}^{\infty}
\phi_{E,\mu,\beta} (\epsilon,n)}
{\biggr \rangle}.
\label{eq:gene_dist_k}
\end{eqnarray}
%%%%%%%%%%%%%%%%%%%%%%%%%%%%%%%%%%%%%%%%%%%%%%%
Now it is 
time for us to 
stress that 
the Ehrenfest or Monkey class 
is recovered 
if 
we choose 
the effective 
Boltzmann factor 
$\phi_{E,\mu,\beta} (\epsilon,n)$ as 
follows \cite{Luck}.
%%%%%%%%%%%%
\begin{eqnarray}
\phi_{E,\mu,\beta} (\epsilon,n) & = & 
\left\{
\begin{array}{ll}
(n!)^{-1}\,{\exp}
\left[
-\beta (E(\epsilon,n)-n\mu)
\right]
 & (\mbox{Ehrenfest class}), \\
{\exp}
\left[
-\beta (E(\epsilon,n)-n\mu)
\right] & (\mbox{Monkey class}).
\end{array}
\right.
\label{eq:Monkey}
\end{eqnarray}
%%%%%%%%%%%%%%%%
It should be noted 
that in our formalism, the distinction between 
two models only comes from 
the difference of the effective Boltzmann factor 
(\ref{eq:Monkey}). 
We here also comment on the effect of the disorder to the dynamics
of urn models. In the uniform case without disorder,
all urns are equivalent, and we do not distinguish each urn
in the Monkey class.
On the other hand, in the disordered urn model,
each urn $i$ has own label $\epsilon_i$,
and hence they are distinguishable at least in principle.
However, we assume that this heterogeneity of urns
does not change the ``Monkey'' nature in the Monkey class;
we cannot see the disorder parameter assigned to each urn from outside,
and the dynamics (so called `box-to-box choice') is not changed.
%%%%%%%%%%%%%%%%%%%%%%%%%%%%%%%%%%%%%%%%%%%%%%%%%%%%%%%%%%%%%%%%%%%%%%%
\section{Ehrenfest class urn model with disorders}
\label{sec:Ehrenfest}
%%%%%%%%%%%%%%%%%%%%%%%%%%%%%%%%%%%%%%%%%%%%%%%%%%%%%%%%%%%%%%%%%%%%%%
As an demonstration 
of the Ehrenfest urn model whose 
thermodynamic properties 
are specified by 
equations (\ref{eq:saddle}) and 
(\ref{eq:dist}), 
we introduce a kind of disordered Ehrenfest urn models 
and consider the condition 
on which the power-law appears.
To this end, we choose the energy function 
$E(\epsilon,n)$ as 
%%%%%%%%%%%%%%%%%%%%%
\begin{eqnarray}
E(\epsilon,n) & = & 
-\epsilon n 
\label{eq:ene9},
\end{eqnarray}
%%%%%%%%%%%%%%%%%%%%%%
where 
$\epsilon$ means an urn-dependent disorder of the 
system,
and we here assume that $\epsilon$ takes a value in the range 
$[0,1]$ randomly, that is, $D(\epsilon)=\Theta (\epsilon) 
-\Theta (\epsilon -1)$ with the step 
function $\Theta (\cdots)$. 
Of course, we might choose 
the other distribution $D(\epsilon)$, 
however, the main issue of this section is 
to examine whether the power-law appears 
in the occupation probability distribution 
when we introduce the disorder, and for this purpose, 
the simplest choice is enough. 
The tendency of this energy function 
to force each urn of the system to gather 
balls as much as possible results in the fact that 
{\it the rich get richer} 
as its collective behavior. 
We should mention that 
in \cite{Leuzzi}, the co-called 
Backgammon model \cite{Ritort} described by the cost function 
$E(\epsilon_{i},n_{i}) = 
-\epsilon_{i} \delta_{n_{i},0}$ for 
each urn was studied. 
In the model, the cost decreases 
if and only if each urn is empty. 
In this sense, the model we deal with 
in this section 
is regarded as an opposite situation 
of our model. Therefore, it is 
interesting to investigate whether 
there exists any significant difference 
or the similarity 
between two models from the view point 
of the occupation probability distribution. 
The extensive studies concerning 
this issue will be our future studies. 

For this choice of the 
energy function (\ref{eq:ene9}), 
the saddle point 
equation (\ref{eq:saddle}) leads to 
%%%%%%%%%%%%%%
\begin{eqnarray}
z_{s} & = & 
\frac{\beta \rho}
{{\rm e}^{\beta}-1}.
\label{eq:saddle2}
\end{eqnarray}
%%%%%%%%%%%%%%
From equation (\ref{eq:dist}), 
the occupation probability 
for the choice (\ref{eq:ene9}), 
$P (k)$, is 
given by 
%%%%%%%%%%%%%%%%%%%%%%%
\begin{eqnarray}
P (k) & = & 
\frac{z_{s}^{k}}{k!}
\int_{0}^{1}
d\epsilon 
\, {\exp}
\left(
\beta \epsilon k - 
z_{s}{\rm e}^{\beta \epsilon}
\right).
\end{eqnarray}
%%%%%%%%%%%%%%%%%%%%%%%%%
In following, we evaluate the above 
occupation probability. 
We first show the phase diagram 
that indicates the area of the power-law behavior. 
%%%%%%%%%%%%%%%%%%%%%%%%%%%%%%%%%%%%%%%%%%
\begin{figure}[ht]
\begin{center}
\includegraphics[width=7.7cm]{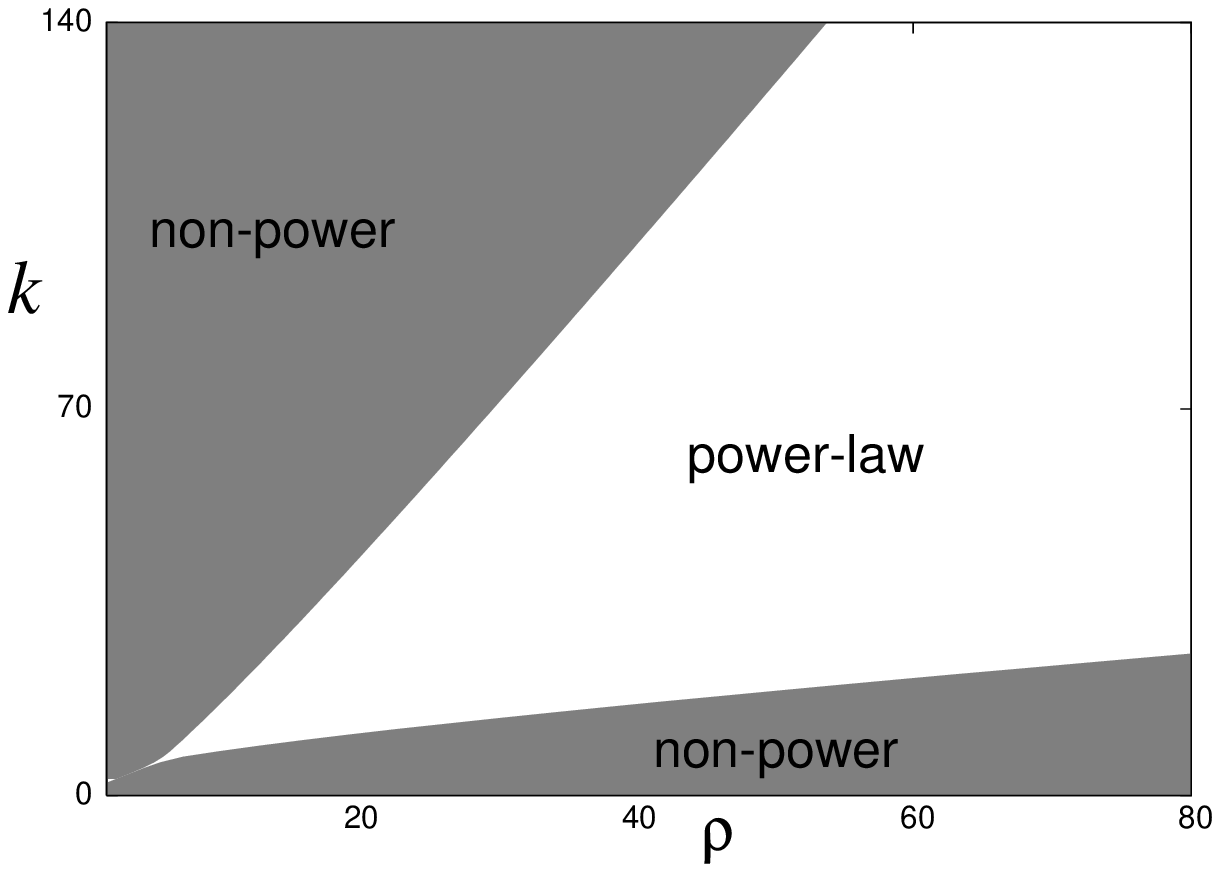}
\includegraphics[width=7.7cm]{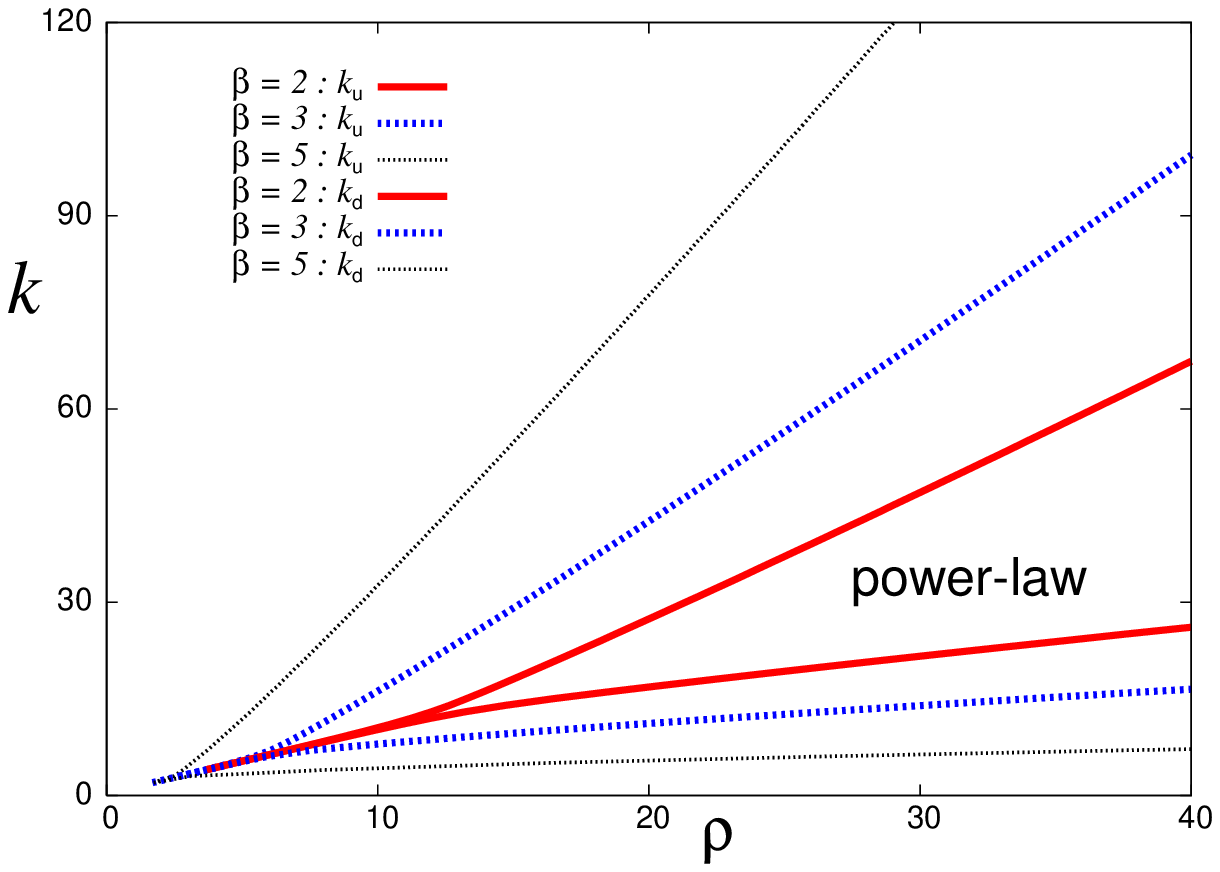}
\end{center}
\caption{\footnotesize 
The $\rho$-$k$ phase diagrams concerning 
the heavy tailedness of the occupation 
probability. 
In the phase diagram 
for $\beta=3$ (left panel), 
the occupation 
probability follows 
non-power, exponential-law. 
The power-law regime 
exists for 
$\rho \geq \rho_{\rm b}=2.697423$. 
In the right panel, 
we plot the phase boundaries 
$k_{d},k_{u}$ as a function of density 
$\rho$ for several values of inverse temperature, 
namely, 
$\beta=2,3$ and $5$.  
}
\label{fig:fgJPA2}
\end{figure}
%%%%%%%%%%%%%%
\mbox{} 
In Figure \ref{fig:fgJPA2} (left), 
we show the phase diagram for the case of 
$\beta=3$. The shaded area in 
this figure means 
non-power, exponential-law region. 
From this figure, we find that 
while the upper bound (the cut-off) 
$k_{\rm u}$ and the lower bound $k_{\rm d}$ 
increase as the density $\rho$ increases,
the cut-off $k_{\rm u}$ increases much more 
quickly than the lower bound $k_{\rm d}$. As the result, 
the heavy tailed power-law $k^{-1}$-region is broadened by increase of 
the density $\rho$. 
In the right panel of 
Figure \ref{fig:fgJPA2}, 
we display 
the inverse temperature 
dependence of 
the area of the power-law. 
As temperature increases, 
the area of the power-law shrinks to zero. 
Then, we have a 
Poisson-law in whole 
region of the phase diagram in the high-temperature 
limit of $\beta =0$. 

We explain the detail of the evaluations as follows. 
We first consider the high-temperature limit 
$\beta=0$. 
For this case, 
the saddle point (\ref{eq:saddle2}) 
leads to $z_{s} = \rho$ and 
we obtain 
%%%%%%%%
\begin{eqnarray}
P (k) & = & 
\frac{\rho^{k}}{k!} \,
{\exp}
\left(
-\rho
\right)
\label{eq:Poisson},
\end{eqnarray}
%%%%%%%%%%
which is nothing but a Poisson distribution. 

\begin{figure}[ht]
\begin{center}
\includegraphics[width=7.7cm]{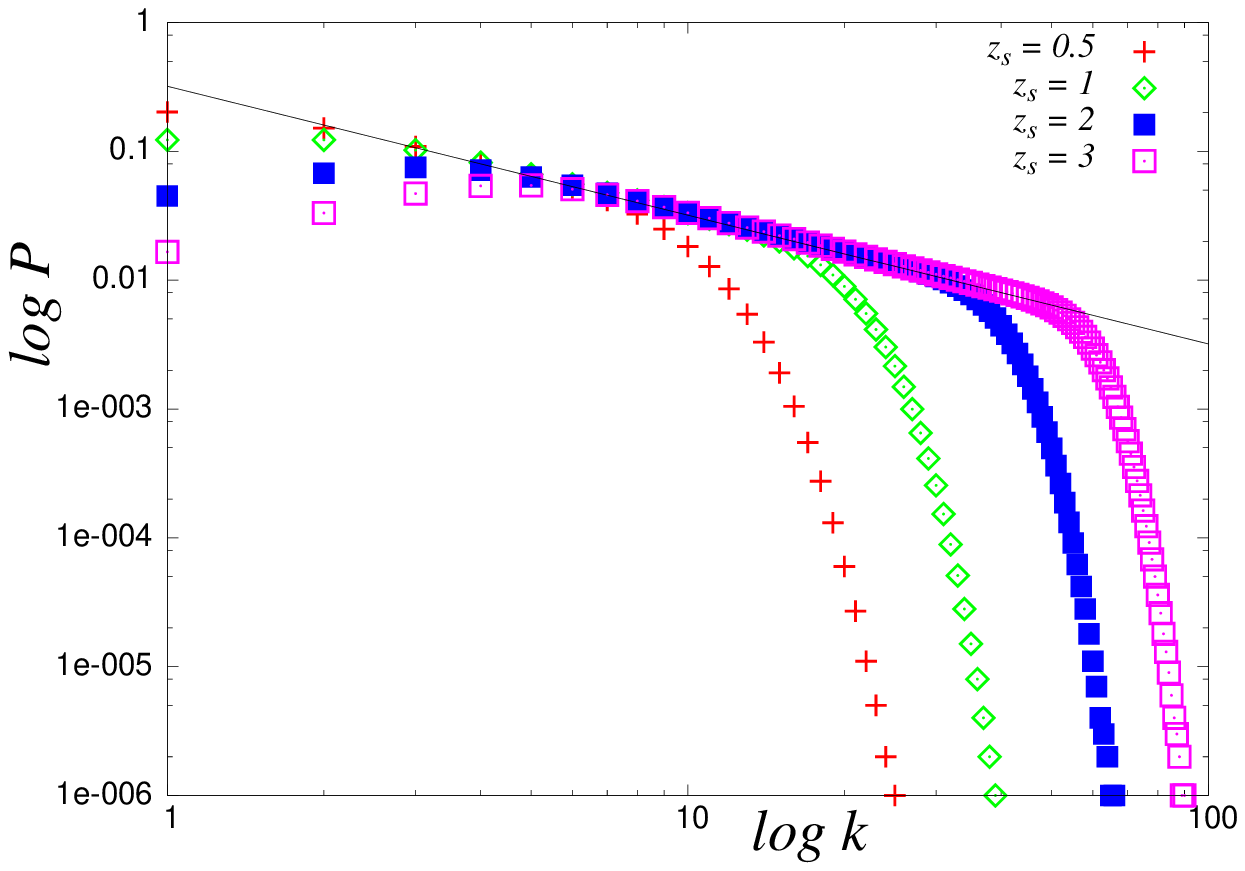}
\includegraphics[width=7.7cm]{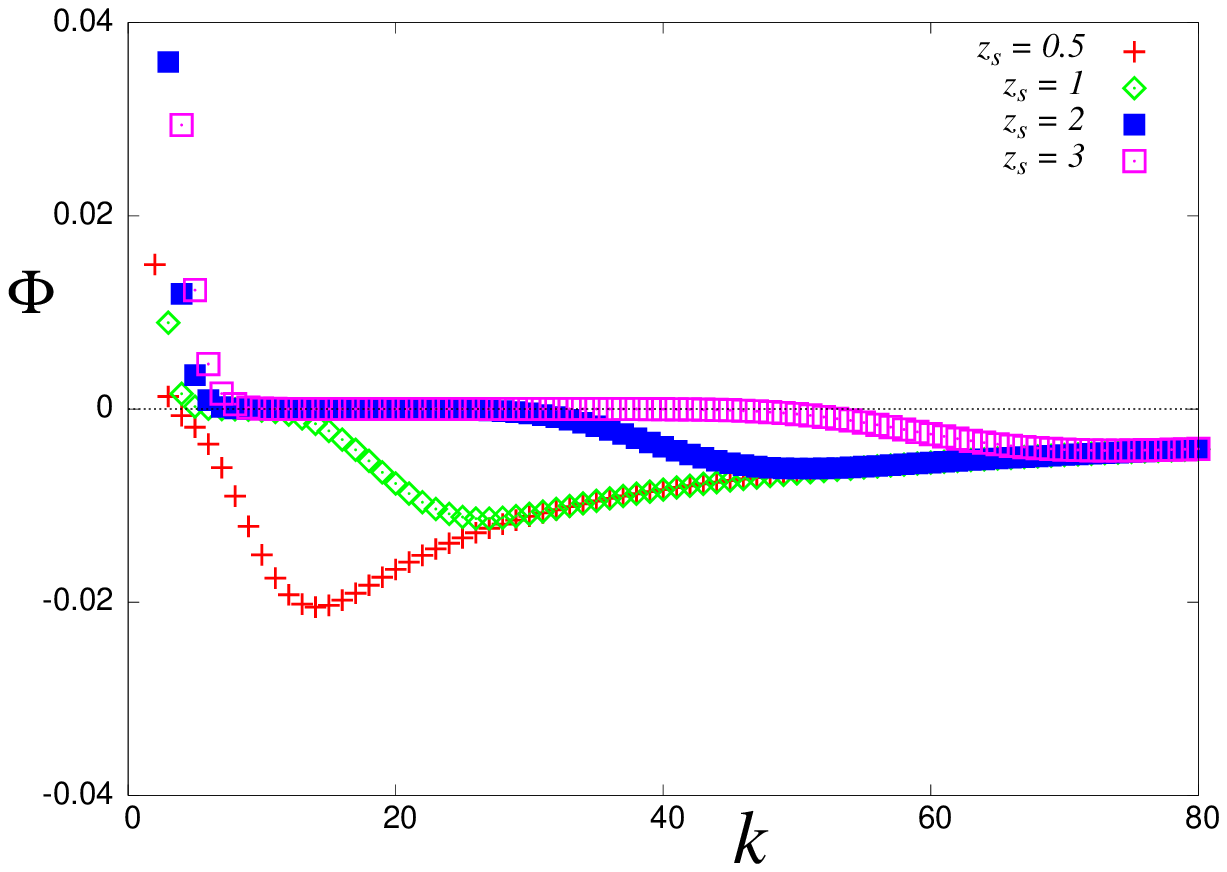}
\end{center}
\caption{\footnotesize 
The Log-Log plot of occupation 
probability $P (k)$ 
as a function of $k$ for several values of 
$z_{s}$, namely, 
$z_{s}=0.5,1,2$ and $z_{s}=3$, 
namely, 
$\rho=3.180923, 6.361846, 
12.723691$ and $19.085537$.
We set the inverse temperature $\beta=3.0$. 
The right panel shows the behavior of $\Phi (k)$. 
}
\label{fig:fgJPA1}
\end{figure}

For finite temperature $\beta > 0$, the occupation 
probability $P (k)$ is rewritten as 
%%%%%%
%%%%%%%%%%%%%%%
\begin{eqnarray}
P (k) & = & 
\frac{1}{\beta k!}
\left[
\Gamma_{k}
\left(
\frac{\beta \rho}{{\rm e}^{\beta}-1}, 
\infty
\right) - 
\Gamma_{k}
\left(
\frac{\beta \rho \,{\rm e}^{\beta}}
{{\rm e}^{\beta}-1}, \infty
\right)
\right]
\label{eq:henkan},
\end{eqnarray}
%%%%%
where we defined 
the incomplete Gamma function by 
$\Gamma_{k}(a,b) = 
\int_{a}^{b}
t^{k-1} 
{\rm e}^{-t}dt$. 
In Figure \ref{fig:fgJPA1}, 
we plot the occupation 
probability $P(k)$ for 
several values of $\rho$ for $\beta=3$,  
$z_{s}=0.5,1,2$ and $3$,  
namely, 
$\rho=3.180923, 6.361846, 
12.723691$ and $19.085537$. 
%%%%%%%%%%%%%%%%%%%%%%%%%%%%%%%%%%%%%%%%%%
%%%%%%%%%%%%%%
From this figure, we find 
that $P (k)$ obeys a power-law 
$\sim k^{-1}$ in the intermediate 
regime of $k$ and there exists a cut-off 
value from which the distribution decays exponentially. 
To see the existence of the cut-off value explicitly, 
we rewrite the distribution $P (k)$ as 
%%%%%%%%%%%%%%%%%
\begin{eqnarray}
\fl
P (k) & = &  
\frac{1}{\beta k}
-\Phi (k), \,\,\,\,\,
\Phi (k) \equiv  
\frac{1}{\beta k!}
\left[
\Gamma_{k}
\left(
0,
\frac{\beta \rho}
{{\rm e}^{\beta \rho}-1}
\right) - 
\Gamma_{k}
\left(
\frac{\beta \rho \, {\rm e}^{\beta}}
{{\rm e}^{\beta}-1},
\infty
\right)
\right].
\end{eqnarray}
%%%%
In Figure \ref{fig:fgJPA1} (right), 
we plot the function 
$\Phi (k)$ for 
several values of 
$z_{s}$ at $\beta =3$. 
%%%%%%%%%%%%%%
We easily find that 
in the range of 
$[k_{\rm d},k_{\rm u}]$ 
defined in terms of the function 
$\Phi (k)$ as 
$k_{\rm d} \equiv \min\{k|\Phi (k)=0\}$, 
$k_{\rm u} \equiv \max\{k|\Phi (k)=0\}$, 
the occupation probability 
$P(k)$ follows 
a power-law distribution $\sim k^{-1}$. 
Thus, we specified 
the region $(\rho,k)$ in which 
a power-law heavy tail 
appears in 
the occupation 
distribution as shown in Figure \ref{fig:fgJPA2}.

As shown in this section, for 
the Ehrenfest class disordered urn model, we could 
clarify the control parameters of the system for which 
the heavy tail power-law emerges in 
the occupation probability. 
In the next section, 
we consider a Monkey class urn model with disorders. 
%%%%%%%%%%%%%%%%%%%%%%%%%%%%%%%%%%%%%%%%%%%%%%%%%%%%%%%%%%%%%%%%%%%%%%%%%%%
\section{Bose-Einstein condensation and emergence of the heavy tail}
\label{sec:BE}
%%%%%%%%%%%%%%%%%%%%%%%%%%%%%%%%%%%%%%%%%%%%%%%%%%%%%%%%%%%%%%%%%%%%%%%%%%
In the previous section, 
we evaluated the asymptotic form of 
the occupation probability $P (k)$ for the urn model of the 
Ehrenfest class with energy function $E(\epsilon,n)=-\epsilon n$. 
Obviously, 
from the view point of the energy cost, 
it is a suitable strategy for each urn to gather 
balls as much as possible. 
In that sense, 
this case should be referred by the concept {\it the rich get richer} 
in the context of social networks. 
However, by using the general definition of the 
problem, we freely choose the energy function 
for both the Ehrenfest and Monkey classes. 

In this section, 
for the Monkey class urn model 
whose 
thermodynamic properties 
are 
defined by equations 
(\ref{eq:gene_saddle}),
(\ref{eq:gene_dist_k}) and 
(\ref{eq:Monkey}), 
we evaluate 
$P (k)$ for 
a specific 
choice of energy function $E(\epsilon,n)$. 
We choose 
the energy $E(\epsilon,n)$ as 
%%%%%%%%%%%%%%%%%%%%
\begin{eqnarray}
E(\epsilon,n) & = & 
\epsilon n \,\,\,\,\,(\epsilon  \geq 0). 
\label{eq:energy_en} 
\end{eqnarray}
%%%%%%%%%%%%%%%%%%%%%%%%%%%%%%
We should notice that 
for this simple choice of 
the energy function, 
the urn labeled by 
$\epsilon \neq 0$ 
is hard to 
gather the balls. 
On the other hand, 
the urn with $\epsilon=0$ 
energy level 
easily gathers the balls. 
The urn model having this type of 
energy function 
does not agree with 
the concept 
{\it the rich get richer}. 
Nevertheless, 
we use the energy function (\ref{eq:energy_en}) 
because 
as we shall see below, a kind of condensation 
with respect to the urns occurs for this choice 
of energy function, and 
as the result, 
the power-law in the tail of 
the occupation probability emerges.

For a given choice of 
$D(\epsilon)$ as 
the density of state, 
namely, 
degeneracy of the energy level of the 
urn, we rewrite 
the saddle point equation 
(\ref{eq:gene_saddle}) as 
follows: 
%%%%%%%%%
\begin{eqnarray}
\rho & = & 
\int_{0}^{\infty}
\frac{D(\epsilon)\, d\epsilon}
{z_{s}^{-1}\,{\rm e}^{\beta \epsilon}-1}
\label{eq:rho_BE1}.
\end{eqnarray}
%%%%%%%%%%%%%
To proceed to the next stage 
of the calculation, we 
choose 
the density of state 
$D(\epsilon)$ explicitly as 
%%%%%%%%%%%%
\begin{eqnarray}
D(\epsilon) & = & 
\varepsilon_{0} \sqrt{\epsilon},
\end{eqnarray}
%%%%%%%%%%%%%%%%%%%
where  
$\varepsilon_{0}$ is a 
constant. 
Although 
we chose the above form, 
more general setup of the argument 
by choosing 
$D(\epsilon)=\varepsilon_{0} \epsilon^{\alpha}$ 
is possible. We shall discuss the result 
for this kind of generalization later on. 
Then, 
the equation 
(\ref{eq:rho_BE1}) is 
rewritten by 
%%%%%%%%%%%%%%%%%%%%%%
\begin{eqnarray}
\rho & = & 
\int_{0}^{\infty}
\frac{\varepsilon_{0} \sqrt{\epsilon} \,d\epsilon}
{z_{s}^{-1}\,{\rm e}^{\beta \epsilon}-1}
+ \rho_{\epsilon=0}
\label{eq:rho_BE2},
\end{eqnarray}
%%%%%%%%%%%%%%
where 
$\rho_{\epsilon=0}$ means 
the density of balls in 
the urn labeled 
by the zero-energy level $\epsilon=0$.
We should notice that 
the second term appearing 
in the right hand side of 
equation (\ref{eq:rho_BE2}), namely, 
$\rho_{\epsilon=0}$ 
vanishes in the thermodynamic limit $N \to \infty$
when a condensation does not occur. 
In other words,
when the condensation arises,
the second term, $\rho_{\epsilon=0}$,
becomes from zero to a finite value;
this means that an urn with $\epsilon = 0$ becomes
to have a macroscopic number of balls.

In following, 
we show the system undergoes 
a condensation and 
investigate the behavior of 
the system when the density $\rho$ 
increases beyond the critical point $\rho_{\rm c}$ 
for a given finite inverse-temperature 
$\beta$. 
%%%%%%
\begin{itemize}
%%%%%%%%%%%%%%%%%%%%%%%%%%%%%%%%%%%
\item
Before condensation: 
$\rho < \rho_{\rm c}$ \\
%%%%%%%%%%%%%%%%%%%%%%%%%%%%%%%%
By a simple transformation 
$\beta \epsilon =x$, 
the equation (\ref{eq:rho_BE2}) 
is rewritten 
in terms of 
the so-called  Appeli function (see e.g. \cite{Morse}) 
$b_{n}(z_{s})$ as 
follows:
%%%%%%%%%%%%%%%%%
%%%%%%%%%%%
\begin{eqnarray}
\rho & = & 
\frac{\varepsilon_{0}\,\sqrt{\pi}}
{2}
\beta^{-3/2}\,
b_{3/2}(z_{s}) 
\label{eq:cond_BE},
\end{eqnarray}
%%%%%%
where the Appeli function (see e.g. \cite{Morse}) 
$b_{n}(z_{s})$ is defined by means of the Gamma function 
$\Gamma(n)$ as 
%%%%%%%%%%%
\begin{eqnarray}
b_{n}(z_{s}) & = & 
\frac{1}{\Gamma(n)}
\int_{0}^{\infty}
\frac{\sqrt{x} \,dx}
{z_{s}^{-1}\,{\rm e}^{x}-1}
\label{eq:saddle_before}.
\end{eqnarray}
%%%%%%%%%%%%%%%%%
We should keep in mind that 
$b_{3/2}(z_{s}) \leq 
b_{3/2}(1)=\zeta (3/2) =2.6...$ 
is satisfied 
($b_{n}(1)=\sum_{k=1}^{\infty}
k^{-n}=\zeta(n)$). 
The solution of 
the saddle point equation (\ref{eq:saddle_before}) 
possesses 
a solution $z_{s} < 1$. 
%%%%%%%%%%%%%%%%%%%%%%%%%%%%%%%%%%%%%%%
\item
At the critical point: $\rho = \rho_{\rm c}$ \\
%%%%%%%%%%%%%%%%%%%%%%%%%%%%%%%%%
The critical point 
at which the condensation occurs 
is determined by the radius of 
convergence for the following partition function: 
%%%%%
\begin{eqnarray}
Z & = & 
\sum_{n=0}^{\infty}
z_{s}^{n} {\rm e}^{-\beta \epsilon n} =  
\sum_{n=0}^{\infty}
({\rm e}^{-\beta \epsilon + 
\log z_{s}})^{n}, 
\end{eqnarray}
%%%%
namely, 
$z_{s}=1$ for $\epsilon=0$ 
gives the critical point. 
Therefore, substituting 
$z_{s}=1$ for a given 
density level $\rho$, 
the critical point  
$\rho_{\rm c}$, above which 
the condensation occurs, 
is obtained by  
%%%%%%%%%
\begin{eqnarray}
\rho_{\rm c} & = & 
\frac{\varepsilon_{0} \sqrt{\pi}}
{2} \beta^{-3/2} \,
b_{3/2}(1).
\end{eqnarray}
%%%%%%%%%%%%%%%%%%%%%%%%%%%%%%%%%%%%%%%%%%%%%%%%%%%%
\item 
After condensation: 
$\rho > \rho_{\rm c}$ \\
%%%%%%%%%%%%%%%%%%%%%%%%%%%%%%%%%%%%%%%%%%%%%%%%%%%%
For $\rho > \rho_{\rm c}$, the 
saddle equation of (\ref{eq:saddle_before}) 
no longer has any solution as $z_{s} <1$. 
Obviously, for the solution $z_{s} >1$, 
the partition function diverges. 
Then, we should bear in mind that 
the term $\rho_{\epsilon=0}$ in (\ref{eq:rho_BE2}), 
which was omitted before the condensation, 
becomes $\mathcal{O}(1)$ object and the saddle 
point equation we should deal with is 
not (\ref{eq:saddle_before}) 
but (\ref{eq:rho_BE2}). 
As the result, the equation (\ref{eq:rho_BE2}) 
has a solution 
$z_{s}=1$ even for $\rho > \rho_{\rm c}$ and 
the number of balls $k_{*}$ in the condensation 
state increases linearly in $\rho$ as 
%%%%%%%%%%%%%
\begin{eqnarray}
k_{*} & = & N (\rho - \rho_{\rm c}), 
\end{eqnarray}
%%%%%%
whereas the number of 
balls in excited states 
reaches $\hat{k} \equiv N \rho_{\rm c}$. 
\end{itemize}
%%%%%%%%%%%%%%%%%%%%%%%
Thus, we obtained 
the saddle point $z_{s}$ for 
a given density and inverse-temperature. 
We found that 
the condensation 
is specified by the solution $z_{s}=1$. 
%%%%%%%%%%%%%%%%%%%%%%%%%%%%%%%%%%%%%%%%

We next investigate 
the density dependence of 
the occupation 
probability through the saddle point. 
For the solution of the saddle point equation 
$z_{s}$, the occupation probability  
at inverse temperature 
$\beta$ is evaluated as follows. 
%%%%%%%%%%%%%%%%%%%%%
\begin{eqnarray}
\fl
P (k) =  
\int_{0}^{\infty}
D(\epsilon) 
d\epsilon 
\left(
\frac{{\rm e}^{-\beta \epsilon k}
z_{s}^{k}}
{\sum_{n=0}^{\infty}
{\rm e}^{-\beta \epsilon n}
z_{s}^{n}}
\right) =  
\frac{z_{s}^{k} 
\varepsilon_{0} 
\Gamma(3/2)}
{\beta^{3/2}} 
k^{-3/2} -
\frac{z_{s}^{k+1}
\varepsilon_{0}
\Gamma(3/2)}
{\beta^{3/2}}
(k+1)^{-3/2}
\label{eq:BE_dist2}
\end{eqnarray}
%%%%%%%%%%%%%%%%%%%%%%%%%%
The above occupation probability is valid for 
an arbitrary integer value of 
$k$ for $k \geq 1$. 
%%%%%%%%%%%%%%%%%%%
%%%%%%%%%%%%%%%%%%%%%%%%%%%%%%%%
\begin{figure}[ht]
\begin{center}
\includegraphics[width=10cm]{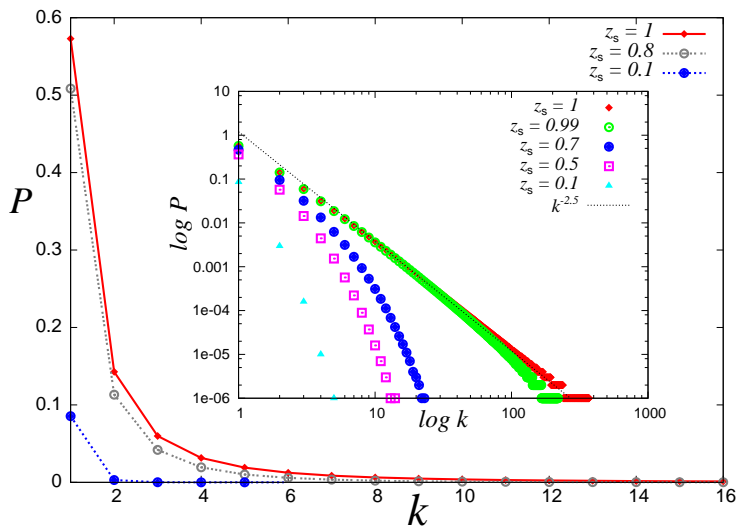}
\end{center}
\caption{\footnotesize 
The behavior of the occupation 
probability (\ref{eq:BE_dist2}) 
in non-asymptotic regime.
We set $\varepsilon_{0}=1$ and $z_{s}$ as 
$z_{s}=0.1,0.8,1.0$, and 
the inverse temperate is 
$\beta=1$. 
The inset of the figure shows 
the asymptotic 
behavior 
of the occupation probability 
$P (k)$ as Log-Log plots for 
the cases of 
$z_{s}=0.1,0.5,0.7,0.99$ and $1$. 
}
\label{fig:fg3}
\end{figure}
%%%%%%%%%%%%%%%%%%%%%%%%%%%%%%%%%%%%%%%
We plot the 
behavior of the occupation probability 
$P (k)$ in 
finite $k$ regime in 
Figure \ref{fig:fg3}. 
In this plot, we set 
$\varepsilon_{0}=1$ and $z_{s}$ as 
$z_{s}=0.1,0.8,1.0$, and 
the inverse temperate is 
$\beta=1$. 
In the inset of the same figure, 
we also show 
the same data in Log-Log scale for 
the asymptotic behavior of 
the probability $P (k)$ 
for several values of $z_{s}$, 
namely, 
$z_{s}=0.1,0.5,0.7,0.99$ and $1$.
%%%%%%%%%%%%%%
From this figure,  
we find that 
the power-law 
$k^{-5/2}$ emerges 
when 
the condensation 
is taken place for $\rho > \rho_{\rm c}$. 
 The numerical 
analysis of 
the occupation probability 
(\ref{eq:BE_dist2}) 
in the limit of $k \to \infty$ 
is easily confirmed 
by asymptotic 
analysis of equation 
(\ref{eq:BE_dist2}). 
We easily find that 
the asymptotic form of 
the wealth distribution 
$P (k)$ behaves as 
%%%%%%%%
%%%%%%%%%%%%%%%%%%%%%%%%%%%
\begin{eqnarray}
\fl 
P (k) =  
\beta^{-3/2}
z_{s}^{k}(1-z_{s})
\varepsilon_{0} 
\Gamma(3/2)
\,k^{-3/2}
+
\frac{3}{2}
\beta^{-3/2}
z_{s}^{k+1}
\varepsilon_{0}
\Gamma (3/2)\, k^{-5/2} + 
{\cal O}
(k^{-7/2}).
\end{eqnarray}
%%%%%%%%%%%%%%
We also should notice that 
a macroscopic number 
of balls $k_{*}$ is gathered to 
a specific urn with 
energy level $\epsilon=0$ when 
the condensation occurs. 
As the result, 
the term such as 
$\sim\, (1/N) \delta (k - k_{*})$ 
should be added to the occupation 
probability. 

Let us 
summarize the results as follows:
%%%%%%%%%%%%%%
\begin{eqnarray}
P (k) & = & 
\left\{
\begin{array}{lc}
\frac{\varepsilon_{0} 
(1-z_{s})}
{\beta^{3/2}}\, k^{-3/2} 
\,{\exp}
\left[
-k\log (1/z_{s})
\right] & (\rho < \rho_{\rm c}: z_{s} <1), \\
%%%%%%%
\frac{3 \varepsilon_{0} 
\Gamma (3/2)}
{2\beta^{3/2}}\, k^{-5/2} + 
\frac{1}{N} 
\delta (k-k_{*}) & 
(\rho \geq \rho_{\rm c}: z_{s}=1).
\end{array}
\right.
\label{eq:result3}
\end{eqnarray}
%%%%%%%%%%%%%%%%%%%%%%%%
The scenario of the condensation 
is as follows.
For a given $\rho < \rho_{\rm c}$,
one can find a solution $z_{s} < 1$ 
for the saddle point equation (\ref{eq:rho_BE2}),
and hence the second term of equation (\ref{eq:rho_BE2}) 
is zero in the thermodynamic limit.
Then, for non-condensed $N\rho$ balls, 
the occupation probability 
follows $\sim \, k^{-3/2}\,{\rm e}^{-k}$-law. 
Namely, urns possessing a large number of 
balls do not appear due to 
the repulsive force as $E=\epsilon n$. 
When $\rho > \rho_{\rm c}$,
the saddle point $z_{s}$ is fixed as $z_{s} = 1$;
if $z_{s} > 1$, the first term of the saddle point 
equation (\ref{eq:rho_BE2})
has a singularity.
Therefore, in order to avoid the singularity,
the second term of the saddle point equation (\ref{eq:rho_BE2}) 
becomes from zero to a finite value.
As the result, 
the occupation probability 
is described by the $k^{-5/2}$-law with 
a delta peak which corresponds to 
an urn of $\epsilon=0$ gathering 
the condensed $N(\rho - \rho_{\rm c})$ balls. 
This corresponds to the condensation phenomena 
in the disordered urn model.
In particular, 
the occurrence of the condensation 
in the disordered urn model treated in the present paper
is characterized 
by the transition from 
the exponential-law to 
the heavy tailed power-law. 
%%%%%%%%%%%%%%%%%%%
We also mention the effect of disorder on 
the power-law behavior of the occupation probability. 
We easily find that the power-law 
behavior disappears 
when one cancels the disorder of the system 
by choosing the density of the energy 
such as $D(\epsilon) = \delta (\epsilon - 
\hat{\epsilon})$ ($\hat{\epsilon}$ is a constant). 
This fact means that the disorder appealing in the system 
possesses a central role 
to make the occupation probability to have 
a power-law behavior. 

We should notice that in the above argument, 
the solution $z_{s}=1$ that indicates the condensation 
does not change even if we choose the density 
as $D(\epsilon)=\varepsilon_{0} \epsilon^{\alpha},\,
(\alpha \geq 0)$. 
For this choice, the $\rho_{\rm c}$ is given by 
%%%%%%
\begin{eqnarray}
\rho_{\rm c} & = & 
\varepsilon_{0} 
\beta^{-1-\alpha}
\int_{0}^{\infty}
\frac{x^{\alpha} dx}
{{\rm e}^{x}-1}.
\end{eqnarray}
%%%%%%%%%%%
Then, one obtains the following 
occupation 
probability 
%%%%%%%%
\begin{eqnarray}
P(k) & = & 
\frac{z_{s}^{k} \varepsilon_{0}
\Gamma (\alpha+1)}
{\beta^{\alpha+1}}
k^{-\alpha-1} - 
\frac{z_{s}^{k+1} \varepsilon_{0} 
\Gamma (\alpha +1)}
{\beta^{\alpha +1}}
(k+1)^{-\alpha -1}. 
\end{eqnarray}
%%%%%
At the end of this section, 
we should mention 
the result for the uniform 
distribution of $\epsilon$, that is, 
the case of $\alpha=0$ leading to 
$D (\epsilon) = \Theta (\epsilon) - 
\Theta (\epsilon-1)$. 
For this choice, we have 
the following occupation 
probability 
%%%%%%%%%%%%%%%
\begin{eqnarray}
P(k) & = & 
\frac{z_{s}^{k}}
{\beta k}
(1-{\rm e}^{-\beta k})
-
\frac{z_{s}^{k+1}}
{\beta (k+1)}
(1-
{\rm e}^{-\beta (k+1)}).
\end{eqnarray}
%%%%%%
Then, beyond the critical density 
$\rho_{\rm c} =
\int_{0}^{1}
d\epsilon
/({\rm e}^{\beta \epsilon} -1)= 
\beta^{-1} 
\sum_{n=1}^{\infty}
(1-{\rm e}^{-\beta n})/n$, 
%%%%%%
the occupation probability $P(k)$ behaves as 
%%%%%
\begin{eqnarray}
P(k) & = &  
\frac{\beta^{-1}}
{k(k+1)}
+ 
\beta^{-1}
\left(
\frac{k{\rm e}^{-\beta} -k-1}
{k(k+1)}
\right) {\rm e}^{-\beta k}.
\end{eqnarray}
%%%%%
Therefore, after the condensation, 
the crossover 
from the $k^{-1}{\rm e}^{-\beta k}$-law to 
the $k^{-2}$-law is observed 
around $k \sim \beta^{-1}$ and as the result, 
the power-law heavy tail appears. 
%%%%%%%%%%%%%%%%%%%%%%%%%%%%%%%%%%%%%%%%%%%%%%%%%%%%%%%%%%%%%
\section{Interpretation from a view point of macro economics}
\label{sec:Econophysics}
%%%%%%%%%%%%%%%%%%%%%%%%%%%%%%%%%%%%%%%%%%%%%%%%%%%%%%%%%%%%%%%
In this section, we reconsider the results obtained in 
the previous sections from 
a view point of macro economics. 
It is easy for us 
to regard the occupation 
probability as wealth distribution 
when we notice 
the relations:  balls - {\it money} and 
urns - {\it people in a society}. 
In following, we attempt to 
find an interpretation of 
the condensation and the emergence 
of the Pareto-law \cite{Pareto} 
in terms of wealth differentials
\cite{Angle,Solomon,Ispolatov,Bouchaud,Dra1,Chatt,Fujiwara,Bikas,Burda}. 

In section \ref{sec:Ehrenfest}, we devoted our analysis to 
extremely large income regimes 
(the tail of the wealth distribution), 
however, it is quite important for us 
to consider the whole range of the wealth. 
As reported in \cite{Bikas}, 
the wealth distribution 
for small income regime 
follows the Gibbs/Log-normal law and 
a kind of transition to the Pareto-law phase is observed. 
For the whole range distribution of 
the wealth, the so-called Lorentz curve 
\cite{Kakwani,Dra2,Silva} is 
obtained. The Lorentz curve is given in terms of the 
relation between the cumulative 
distribution of wealth and 
the fraction of 
the total wealth. 
Then, the so-called Gini index \cite{Kakwani,Dra2,Silva,SazukaInoue}, 
which is a traditional, popular and one of the most basic 
measures for wealth differentials, 
could be calculated. 
The index could be changed from 
$0$ (no differentials) to 
$1$ (the largest differentials). 
For the energy function (\ref{eq:energy_en}) 
in the previous section, 
we derived the wealth distribution 
for the whole range of incomes $k$. 
In this section, we evaluate the Gini 
index analytically. 

As we mentioned above, the Lorentz curve 
is determined by the relation between 
the cumulative 
distribution of wealth $X(t)= 
\int_{t_{\rm min}}^{t}
P(k)dk$ 
and 
the fraction of 
the total wealth $Y(t)=\int_{t_{\rm min}}^{t}
kP(k)dk/\int_{t_{\rm min}}^{\infty}
kP(k)dk$ 
for a given wealth distribution $P(k)$. 
For instance, the 
Lorentz curve for the exponential distribution 
$P(k)=\gamma \,{\rm e}^{-\gamma k}$ is 
given by 
%%%%%%%%%%%%
\begin{eqnarray}
Y & = & 
X + (1-X) \log (1-X).
\label{eq:L_exp}
\end{eqnarray} 
%%%%%%%
We should notice that 
the Lorentz curve for the exponential 
distribution is independent of $\gamma$. 

\begin{figure}[ht]
\begin{center}
\includegraphics[width=7.7cm]{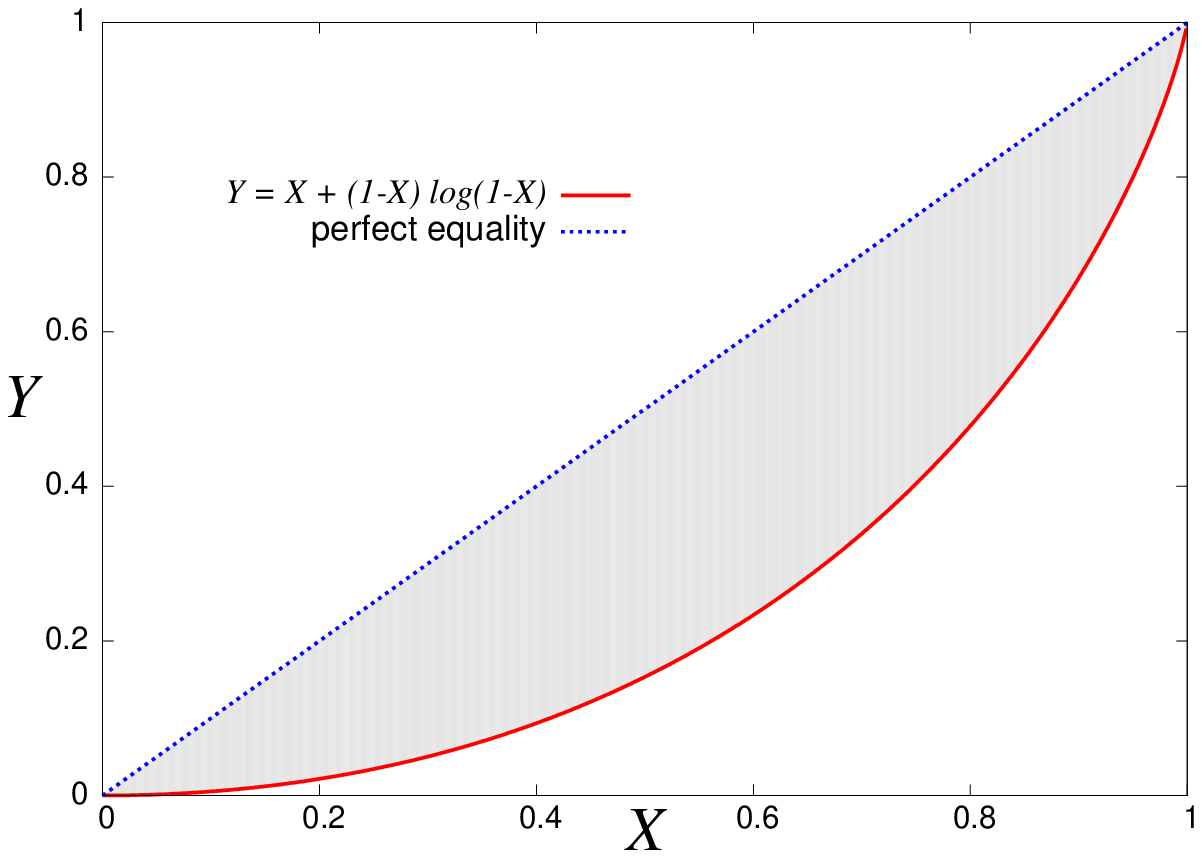}
\includegraphics[width=7.7cm]{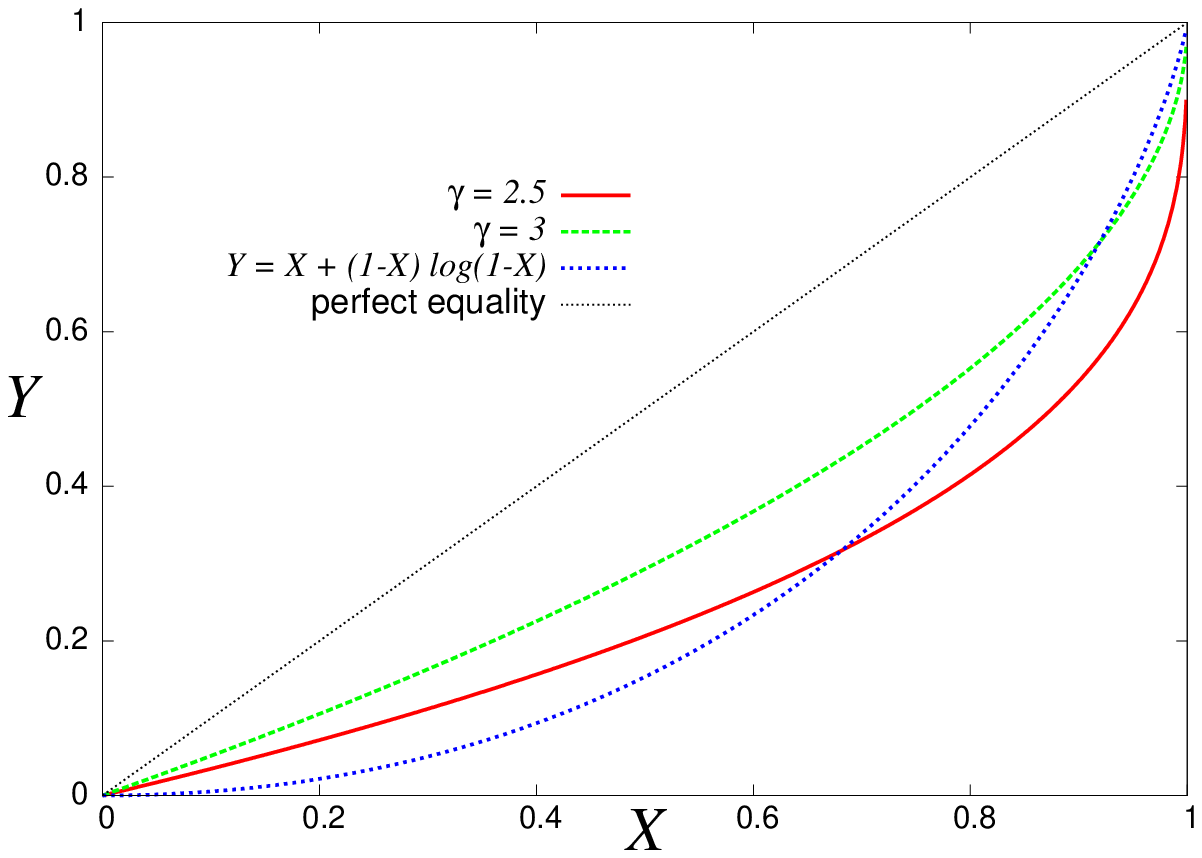}
\end{center}
\caption{\footnotesize 
The left panel: 
The Gini index is obtained 
as an area between 
the perfect equality line 
$Y=X$ and the Lorentz curve 
(e.g. $Y=X+(1-X)\log (1-X)$ for 
exponential distributions). 
The right panel shows the Lorentz curves 
for the exponential 
distribution (\ref{eq:L_exp}) and 
the power-law distributing 
(\ref{eq:L_pow}) 
with several values of $\gamma=2.5$ and $3$. 
}
\label{fig:fg4}
\end{figure}

For the power-law distribution 
$P(k)=(\gamma-1)\,k^{-\gamma} \,\,(\gamma >1)$, 
we have 
%%%%%%%%%%%%%%%%%%%
\begin{eqnarray}
Y & = & 
1- (1-X)^{\frac{\gamma-2}{\gamma-1}}.
\label{eq:L_pow}
\end{eqnarray}
%%%%%%%%%%
as the Lorentz curve. 
This curve depends on the exponent $\gamma$. 
In Figure \ref{fig:fg4}, 
we plot the Lorentz curve 
for the exponential 
distribution (\ref{eq:L_exp}) and 
the power-law distribution 
(\ref{eq:L_pow}) 
with several values of $\gamma$. 
%%%%%%%%%%%%%%%%%%%%%%%%%%%%%%%%
%%%%%%%%%%%%%%%%%%%%%%%%%%%%%%%%%%%%%%%
\mbox{}

\begin{figure}[ht]
\begin{center}
\includegraphics[width=7.7cm]{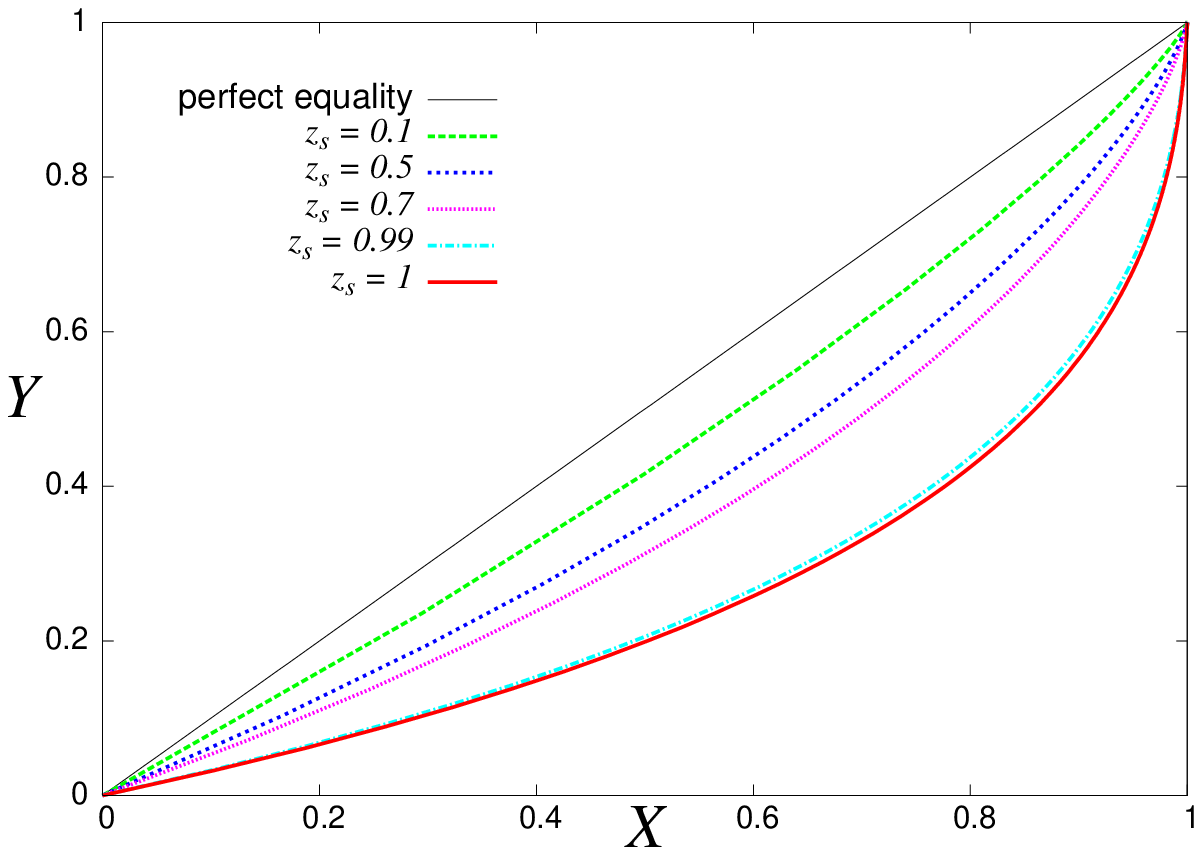}
\includegraphics[width=7.7cm]{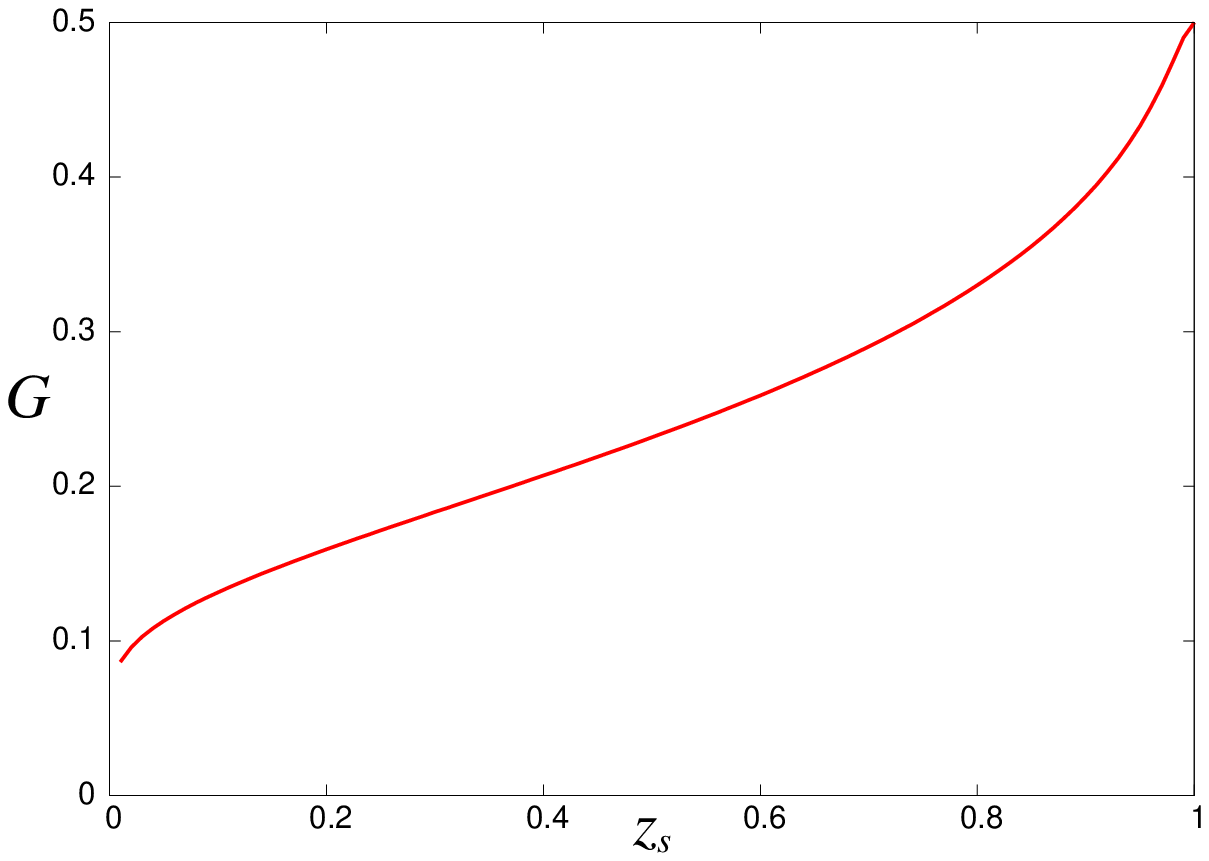}
\end{center}
\caption{\footnotesize 
The left panel: 
the Lorentz curve 
for (\ref{eq:BE_dist2}). 
The right panel shows the 
Gini index for 
several values of 
$z_{s}$.  
}
\label{fig:fg5}
\end{figure}

Then, as shown in the left panel of 
Figure \ref{fig:fg4}, the Gini index $G$ 
is defined as an area 
between the perfect equality line $Y=X$ and 
the Lorentz curve. 
This quantity explicitly reads 
%%%%%%%%%%%%%%%%%%%%%
\begin{eqnarray}
G & = & 
2\int_{0}^{1} (X-Y)dX = 
2 \int_{t_{\rm min}}^{\infty}
(X(t)-Y(t)) \cdot 
\frac{dX}{dt} dt 
\end{eqnarray}
%%%%%%%%%%%%%%
and we have $G=1/2$  \cite{Dra2,Silva} for 
the exponential 
distribution and 
$G=1/(2\gamma-3)$ for 
the power-law distribution. 
As the occupation 
probability 
distribution (\ref{eq:BE_dist2}) 
is defined for $k > 1$, one can 
evaluate the Gini index as 
a function of the saddle point 
$z_{s}$. 
In Figure \ref{fig:fg5}, we plot 
the Lorentz curve (left) for 
several values of $z_{s}$. 
In the right panel, 
the Gini index $G(z_{s})$ is shown. 
We find that the index approaches to 
$1/2$ as $z_{s} \to 1$. 

From the argument in the previous section,  
we easily find that the occupation 
distribution for 
$N \rho_{\rm c}$ non-condensed balls 
beyond 
the critical point 
is modified such as 
$\sim\, k^{-(\alpha +2)}$ by 
choosing the density of the 
energy $D(\epsilon)=\varepsilon_{0} \,
\epsilon^{\alpha}$.
Namely, for the Pareto-law distribution 
$\sim k^{-(\alpha+2)}$, 
the Gini index leads to 
$G=1/(2\alpha+1)$. 
Therefore,  
the condensation 
is specified by the change of the 
Gini index from 
$G=1/2$ to $1/(2\alpha+1)$. 
However, 
we should keep in mind that 
the Gini index 
itself has less information about the 
differentials than the 
wealth distribution. 
For example, 
the Gini index 
for $\alpha=1/2$ of the 
Pareto-law $P(k) \sim 
k^{-(\alpha +2)}$ 
gives the same Gini index 
as the exponential 
distribution. 
This fact stems from 
the definition of 
the Gini index $G$, that is, 
$G$ is defined as an area between 
$Y=X$ and the Lorentz curve. 
It could be possible to 
draw lots of the Lorentz curves 
that give the same area 
(the same Gini index). 
As we explained above, 
it should be noted that 
actually the Gini index is 
one of the measure for 
the earning differentials, however, 
the wealth distribution  
is much more informative than the Gini index. 
%%%%%%%%%%%%%%%%%%%%
%%%%%%%%%%%%%%%%%%%%%%%%%%%%%%%%
%%%%%%%%%%%%%%%%%%%%%%%%%%%%%%%%%%%%%%%
Although the Gini index is less informative 
than the distribution, however, for a given real (empirical) 
data $x_{1} \leq x_{2} \leq \cdots \leq x_{N}$ 
($x$ denotes the amount of money for example. 
Such empirical data are not 
always massive enough for us to specify the distribution), 
one can evaluate it as a statistics by 
$G = (1/N^{2}\mu)
\sum_{i=1}^{N}
(2i-N-1)x_{i}$ with $\mu N=\sum_{r=1}^{N}x_{r}$ \cite{SazukaInoue}. 
Therefore, it is helpful for us to use the Gini index to compare 
the earning differentials between different countries 
(the population of each country should be different and of course 
it is finite $N < \infty$). 
In our analysis of this paper, the distribution 
was analytically obtained in the thermo-dynamic limit because 
we treated an ideal case as a society. 
Nevertheless, even if we encounter more realistic situation 
for which the analytical evaluation of wealth distribution 
is very tough, one can evaluate the earning differentials 
via the Gini index by computer simulations for 
finite population $N$. 
Then, one can 
investigate the earning differentials by comparing 
the numerical results with 
the analytical expressions obtained in this paper. 
%%%%%%%%%%%%%%%%%%%%%%%%%%%%%%%%%%%%%%%%%%%%%%%%%%%
\section{Summary}
\label{sec:Summary}
%%%%%%%%%%%%%%%%%%%%%%%%%%%%%%%%%%%%%%%%%%%%%%%%%%%%%%%
In this paper, we investigated 
equilibrium properties of 
disordered urn model and discuss the condition on 
which the heavy tailed 
power-law appears in 
the occupation probability 
by using statistical physics of disordered 
spin systems. 
We applied our formalism to 
two urn models of 
both the Ehrenfest and 
Monkey classes. 
In particular, for the choice of the energy function 
as $E(\epsilon, n) = \epsilon n$ 
with density of state  
$D(\epsilon) = \varepsilon_{0} \epsilon^{\alpha}$ for 
the Monkey class urn model, 
we found that 
above the critical density 
$\rho > \rho_{\rm c}$ for a temperature, 
the condensation phenomenon
is taken place, and most of the balls falls in an urn with
the lowest energy level. 
As the result, the occupation 
probability  
changes its scaling behavior 
from the 
exponential $k^{-(\alpha +1)}\,{\rm e}^{-k}$-law to 
the $k^{-(\alpha +2)}$ power-law 
in large $k$ regime. 
We also provided 
a possible link 
between our results and 
macro economy, in particular, 
wealth differentials. 

Of course, there might exist the other 
urn models showing the power-law behavior 
after the condensation. 
In fact, we find such 
a case in a recent study on the Ehrenfest 
urn model \cite{Ohkubo2007_2}, in which the occupation 
probability follows a Poisson-law when 
the condensation occurs. 
%%%%%%%%%%%%%%%%%%%%%
Although we provided a piece of 
evidence to show that 
the power-law behavior in the occupation 
probability distribution 
takes place after the condensation 
for several restricted cases of the cost function, 
it is not yet clear whether 
the condensation always causes the power-law or not. 
The nature of the link between them will be 
a central problem to be clarified in future. 
Therefore, as one of our future studies, 
it might be important to investigate 
the universality 
class of urn models that 
shows the power-law behavior in the 
occupation probability beyond the critical point.

We hope that various versions and extensions of the disordered 
urn model, including Backgammon model \cite{Ritort,Leuzzi}, 
could be applied to research area beyond 
conventional statistical physics.

%%%%%%%%%%%%%%%%%%%%%%%%%%%%%%%%%%%%%%%%%%%%%%%%%%%%%
\section*{Acknowledgments}
%%%%%%%%%%%%%%%%%%%%%%%%%%%%%%%%%%%%%%%%%%%%%%%%%%%%%%

One of the authors (J.I.) 
was financially supported 
by {\it Grant-in-Aid 
Scientific Research on Priority Areas 
``Deepening and Expansion of Statistical 
Mechanical Informatics (DEX-SMI)" 
of The Ministry of Education, Culture, 
Sports, Science and Technology (MEXT)} 
No. 18079001. 
We would like to thank 
Enrico Scalas for introducing 
us their very recent studies \cite{Scalas} concerning 
the Ehrenfest urn model.

\end{document}